\documentstyle[12pt]{article}

\author{L.S.F. Olavo\\
Departamento de Fisica ,Universidade de Brasilia - UnB\\
70910-900, Brasilia - D.F. - Brazil}
\title{Quantum Mechanics as a Classical Theory\\
XIV: Connection with Stochastic Processes
}
\date{september, 27, 1996
}

\begin{document}

\maketitle
\begin{abstract}
In this paper we are interested in unraveling the mathematical connections
between the stochastic derivation of Schr\"odinger equation and ours. It
will be shown that these connections are given by means of the time-energy
dispersion relation and will allow us to interpret this relation on more
sounded grounds. We also discuss the underlying epistemology.
\end{abstract}

\section{Introduction}

The search for a stochastic support of quantum mechanics is already known
since the 50s\cite{1} and was a fertile research field in the decades of 50
and 60\cite{2}-\cite{10}.

It is still a sounded field for the investigation of the mathematical and
epistemological foundations of quantum mechanics.

This approach can be seen as the mathematical demonstration that one may
derive the quantum mechanical formalism (Schr\"odinger equation) using only
the formal apparatus of classical statistic mechanics, together with the
Brownian movement theory\cite{Nelson},

In this case the cinematic description of the Brownian movement is the one
related with a movement with no friction---the same used in the
Einstein-Smoluchovski theory\cite{12,13}.

The picture drawn from this approach is such that a particle, submitted to
an external force, remains in dynamic equilibrium because of the balance of
this force with a stochastic force responsible for the random movement.

The important point here is that in a such a theory, where {\bf x}(t) is
considered a stochastic process, it is not possible to define a total time
derivative $d/dt$, since the movement is discontinuous, and we have to
search for substitutes of this operator that might be used to formulate
another {\it new} `Newtonian' theory, formally equivalent to the
mathematical structure of quantum mechanics, as given by the Schr\"odinger
equation. This is amply known in the literature and will be also shown in
the present paper.

We have already shown in some previous papers\cite{eu1}-\cite{eu13} that it
is possible to derive the quantum formalism (Schr\"odinger equation) from
three classical postulates: the first one is the general validity of
Newton's laws for the individual systems, the second is the validity of
Liouville's equation for the {\it ensembles }and the last one is the
possibility of connecting the joint probability density function on
phase-space $F({\bf x},{\bf p};t)$ to a characteristic function $\rho ({\bf x%
},\delta {\bf x};t)$ by means of an infinitesimal transformation 
\begin{equation}
\label{1}\rho ({\bf x},\delta {\bf x};t)=\int e^{i{\bf p}\cdot \delta {\bf x}%
/\hbar }F({\bf x},{\bf p};t)d{\bf p},
\end{equation}
where $\delta {\bf x}$ is considered an infinitesimal displacement.

With these two derivation methods, it would be of some relevance to take a
look at their epistemological and mathematical connection. Indeed, from the
fact that it was already shown that quantum mechanics may be understood as a
stochastic process, it would be rather interesting to know in what point of
our approach this stochastic character was effectively introduced and what
is its ontological status, according to our approach.

This is the objective of the present paper.

To achieve this goal we will develop briefly, in the second section, the
derivation of the Schr\"odinger equation from the stochastic point of view.
We will follow the pioneering papers of Nelson\cite{Nelson}, Kershaw\cite
{Kershaw} and De La Pe\~na\cite{Pena1}, and the revision paper of De La
Pe\~na\cite{Pena2} as our guides in the development of the related formalism
and interpretation.

In the third section, we will show again and very schematically our
derivation of the Schr\"ondiger equation using the infinitesimal
Wigner-Moyal transformation given by (\ref{1}).

In the fourth section we will show how the stochastic derivation is
connected with the one made in the third section.

The last section is left to our final considerations.

\section{Stochastic Derivation}

As we have seen in the previous section, ${\bf x}(t)$ is a stochastic
process and we cannot define a time derivative $d/dt$ for it. This means
that the velocity related with this process cannot be obtained by its direct
derivation, for ${\bf x}(t)$ is not, in general, differentiable.

In this case we have to introduce a finite time interval $\Delta t$, small
compared with the characteristic times of the systematic movement (the one
related with Newton's equation), but large enough compared with the
correlation time of the fluctuating force\cite{Pena2}.

Using this finite time interval we may define the forward time derivative%
\cite{Nelson} as 
\begin{equation}
\label{2}D{\bf x}(t)=\lim _{\Delta t\rightarrow 0^{+}}E_t\frac{{\bf x}%
(t+\Delta t)-{\bf x}(t)}{\Delta t}=\langle \frac{\delta {\bf x}(\Delta t)}{%
\Delta t}\rangle _t,
\end{equation}
where the mean $E\_t$ or $\langle \rangle _t$ is taken over the $\Delta t$
distribution, which means that it is the conditional mean in the interval $%
\Delta t$ and reflects a statistical distribution of the displacements $%
\delta {\bf x}$ \cite{Kershaw}.We may also define the backward derivative as 
\begin{equation}
\label{3}D_{*}{\bf x}(t)=\lim _{\Delta t\rightarrow 0^{+}}E_t\frac{{\bf x}%
(t)-{\bf x}(t-\Delta t)}{\Delta t}=\langle \frac{\delta {\bf x}(-\Delta t)}{%
-\Delta t}\rangle _t,
\end{equation}
where, in general, $D{\bf x}(t)\ne D_{*}{\bf x}(t)$.

Since $\Delta t$ is a very small time interval, we may write the following
expansion 
$$
\frac 1{\Delta t}[f({\bf x}(t+\Delta t),t+\Delta t)-f({\bf x}(t),t)]\approx
\left[ \frac \partial {\partial t}+\frac 1{\Delta t}\sum_i[x_i(t+\Delta
t)-x_i(t)]\frac \partial {\partial x_i}+\right.  
$$
\begin{equation}
\label{4}\left. +\frac 1{2\Delta t}\sum_{ij}[x_i(t+\Delta
t)-x_i(t)][x_j(t+\Delta t)-x_j(t)]\frac{\partial ^2}{\partial x_ix_j}\right]
.
\end{equation}
Taking the mean $\langle \rangle _t$ in the last expression and also taking
the limit $\Delta t\rightarrow 0$ we have (up to second order) 
\begin{equation}
\label{5}Df({\bf x},t)=\left( \frac \partial {\partial t}+{\bf c}\cdot
\nabla +\nu \nabla ^2\right) f({\bf x},t),
\end{equation}
where 
\begin{equation}
\label{6}{\bf c}=\lim _{\Delta t\rightarrow 0}E_t\left[ \frac{{\bf x}%
(t+\Delta t)-{\bf x}(t)}{\Delta t}\right] =D{\bf x}(t),
\end{equation}
and 
\begin{equation}
\label{7}\nu =\lim _{\Delta t\rightarrow 0}E_t\left[ \frac{({\bf x}(t+\Delta
t)-{\bf x}(t))^2}{2\Delta t}\right] .
\end{equation}

We may now split the velocity {\bf c} into two components: the systematic
component ${\bf v}$ and the stochastic one ${\bf u}$ 
\begin{equation}
\label{8}{\bf c}={\bf v}+{\bf u}.
\end{equation}

It is also interesting to define the time-inversion operator $\hat{T}$,
since it will be important to know the behavior of the processes ${\bf v}$
and ${\bf u}$ under its action when making the stochastic modification of
Newton's laws. Hence, we have 
\begin{equation}
\label{9} {\bf c}_*=\hat{T}{\bf c}=\hat{T}D{\bf x}(t)=D_*{\bf x}(t), 
\end{equation}
as in the definition (\ref{3}).

For a general function we have 
\begin{equation}
\label{10} D_*f({\bf x},t)=\left[-\frac{\partial}{\partial t}+{\bf c}_*\cdot
\nabla+\nu_*\nabla^2\right]f({\bf x},t). 
\end{equation}
The two derivative operators are given by 
\begin{equation}
\label{11} \left\{ 
\begin{array}{l}
D=\partial/\partial t+ 
{\bf c}\cdot\nabla+\nu\nabla^2, \\ D_*=-\partial/\partial t+{\bf c}%
_*\cdot\nabla+\nu_*\nabla^2 
\end{array}
\right. . 
\end{equation}

We now define the two operators 
\begin{equation}
\label{12} D_c=\frac{1}{2}(D-D_*) \mbox{ , } D_s=\frac{1}{2}(D+D_*), 
\end{equation}
named `systematic' and `stochastic' derivative, respectively. From relations
(\ref{11}) and (\ref{12}) we have 
\begin{equation}
\label{13} \left\{ 
\begin{array}{l}
D_c=\partial/\partial t+ 
{\bf v}\cdot\nabla-\nu_-\nabla^2, \\ D_s={\bf u}\cdot\nabla+\nu_+\nabla^2 
\end{array}
\right. , 
\end{equation}
where 
\begin{equation}
\label{14} \nu_+=\frac{1}{2}(\nu_*+\nu) \mbox{ , } \nu_-=\frac{1}{2}%
(\nu_*-\nu), 
\end{equation}
and 
\begin{equation}
\label{15} {\bf v}=\frac{1}{2}({\bf c}+{\bf c}_*) \mbox{ , } {\bf u}=\frac{1 
}{2}({\bf c}-{\bf c}_*). 
\end{equation}

In the Newtonian limit $\Delta t\rightarrow 0$ the operator $D_s\equiv 0$
while the operator $D_c$ becomes 
\begin{equation}
\label{16}D_c\rightarrow \frac \partial {\partial t}+{\bf v}\cdot \nabla
=\frac d{dt}, 
\end{equation}
and we may identify ${\bf v}$ with the systematic velocity. In this same
limit ${\bf u}\rightarrow 0$.

To build a dynamic theory we have now to postulate the following relation
between the stochastic acceleration and the stochastic derivative of ${\bf c}
$ 
\begin{equation}
\label{17}m{\bf a}=D{\bf c},
\end{equation}
as a substitute for the equation giving Newton's second law in the Newtonian
limit. Indeed, in this limit 
\begin{equation}
\label{18}D=D_c+D_s\rightarrow \frac d{dt}\mbox{  and  }{\bf c}\rightarrow 
{\bf v}.
\end{equation}
By means of relation (\ref{17}) we have 
\begin{equation}
\label{19}{\bf a}=D_c{\bf v}+D_s{\bf u}+D_c{\bf u}+D_s{\bf v}.
\end{equation}

If the forces involved do not depend upon the velocity, then these forces
are $\hat T$-invariant 
\begin{equation}
\label{20}\hat T{\bf f}={\bf f}.
\end{equation}
But the acceleration ${\bf a}$, as defined in (\ref{19}), is not $\hat T$%
-invariant. Indeed 
\begin{equation}
\label{21}D_c=-\hat TD_c\mbox{ , }D_s=+\hat TD_s;
\end{equation}
thus 
\begin{equation}
\label{22}\hat T{\bf a}=D_c{\bf v}+D_s{\bf u}-D_c{\bf u}-D_s{\bf v}.
\end{equation}

To have ${\bf a}=\hat T{\bf a}$ as required by (\ref{20}) we must have 
\begin{equation}
\label{23}\left\{ 
\begin{array}{l}
D_c
{\bf v}+D_s{\bf u}={\bf a} \\ D_s{\bf v}+D_c{\bf u}=0
\end{array}
\right. .
\end{equation}
Writing 
\begin{equation}
\label{24}{\bf a}={\bf a}_c+{\bf a}_s=D_c{\bf v}+D_s{\bf u},
\end{equation}
then 
\begin{equation}
\label{25}\left\{ 
\begin{array}{l}
{\bf a}_c=D_c{\bf v}=D_c^2{\bf x}(t)=\hat T{\bf a}_c \\ {\bf a}_s=D_s{\bf u}%
=D_s^2{\bf x}(t)=\hat T{\bf a}_s
\end{array}
\right. ,
\end{equation}
or else, by means of the operator $D$ and $D_{*}$, 
\begin{equation}
\label{26}\left\{ 
\begin{array}{l}
{\bf a}_c=\frac 14(D-D_{*})^2{\bf x}(t) \\ {\bf a}_s=\frac 14(D+D_{*})^2{\bf %
x}(t)
\end{array}
\right. .
\end{equation}

Now let ${\bf f}_0$ be an external applied force. This force also has a
stochastic component and may be equally written as a linear combination of
systematic plus stochastic forces. In this case we may write 
\begin{equation}
\label{27}{\bf f}_0=m\lambda _1{\bf a}_c+m\lambda _2{\bf a}_s, 
\end{equation}
where $\lambda _1$ and $\lambda _2$ are constants.

In the newtonian limit we must have ${\bf f}_0=m{\bf a}_c$ which implies
that $\lambda =1$. Now putting $\lambda _2=-\lambda $ we get 
\begin{equation}
\label{28}{\bf f}_0=m({\bf a}_c-\lambda {\bf a}_s). 
\end{equation}
Since the total force is 
\begin{equation}
\label{29}{\bf f}=m({\bf a}_c+{\bf a}_s), 
\end{equation}
we finally get 
\begin{equation}
\label{30}{\bf f}={\bf f}_0+m(1+\lambda ){\bf a}_s, 
\end{equation}
or 
\begin{equation}
\label{31}\left\{ 
\begin{array}{l}
{\bf f}_0=D_c{\bf v}-\lambda D_s{\bf u} \\ D_c{\bf u}+D_s{\bf v}=0 
\end{array}
\right. , 
\end{equation}
which may be written explicitly as 
\begin{equation}
\label{32}\left\{ 
\begin{array}{l}
\partial 
{\bf v}/\partial t+({\bf v}\cdot \nabla ){\bf v}-\nu _{-}\nabla ^2{\bf v}%
-\lambda ({\bf u}\cdot \nabla ){\bf u}-\lambda \nu _{+}\nabla ^2{\bf u}={\bf %
f}_0 \\ \partial {\bf u}/\partial t+({\bf v}\cdot \nabla ){\bf v}+{\bf u}%
\cdot \nabla ){\bf v}+\nu _{+}\nabla ^2{\bf v}-\nu _{-}\nabla {\bf u}=0 
\end{array}
\right. . 
\end{equation}

In the newtonian limit\cite{Pena2} we have 
\begin{equation}
\label{33} \nu_+=\nu_-=0 \rightarrow {\bf u}=0 
\end{equation}
thus giving 
\begin{equation}
\label{34} m\frac{d{\bf v}}{dt}={\bf f}_0, 
\end{equation}
as desired.

Assuming that $\nu _{+}$ and $\nu _{-}$ will depend only upon the time, the
velocities are rotational-free and that the external forces are derivable
from a potential $V$, we may rewrite the system of equations (\ref{32}) as 
\begin{equation}
\label{35}\left\{ 
\begin{array}{l}
\partial 
{\bf v}/{\partial t}+\nabla [{\bf v}^2/2-\nu _{-}\nabla \cdot {\bf v}%
-\lambda {\bf u}^2/2-\lambda \nu _{+}\nabla \cdot {\bf u}]=-\nabla V/m \\ {%
\partial {\bf u}}/{\partial t}+\nabla [{\bf v}\cdot {\bf u}+\nu _{+}\nabla
\cdot {\bf v}-\nu _{-}\nabla \cdot {\bf u}]=0
\end{array}
\right. .
\end{equation}

To obtain the Schr\"odinger equation from equation (\ref{35}) we have only
to make the {\it ansatz} 
\begin{equation}
\label{36}{\bf v}=2D_0\nabla S\mbox{  and  }{\bf u}=2D_0\nabla R, 
\end{equation}
with $\nu _{+}=D_0$, $\nu _{-}=0$ and $\psi _{\pm }=exp(R\pm iS/\sqrt{%
-\lambda })$, where $R$ and $S$ are functions depending upon ${\bf x}(t)$
and $t$. After some algebra we get 
\begin{equation}
\label{37}\mp 2imD_0\sqrt{-\lambda }\frac{\partial \psi _{\pm }}{\partial t}%
=-2m\lambda D_0^2\nabla ^2\psi _{\pm }+V\psi _{\pm }, 
\end{equation}
where $V$, as was said above, is the potential function related with the
external force ${\bf f}_0$.

Since the parameters $\lambda $ and $D_0$ appear in equation (\ref{37}) only
through the product $D_0\sqrt{-\lambda }$, it is clear that we may adjust
the scale through $D_0$ and take $|\lambda |=1$\cite{Pena2}.

If $\lambda =-1$ we get the equation 
\begin{equation}
\label{38}\mp 2mD_0\partial \psi _{\pm }/\partial t=2mD_0^2\nabla ^2\psi
_{\pm }+V\psi _{\pm }
\end{equation}
having as its solution 
\begin{equation}
\label{39}\psi _{\pm }=e^{R\pm S}=\rho ^{1/2}e^{\pm S}\mbox{ onde }\rho
=\psi _{+}\psi _{-}.
\end{equation}
This equation is of the parabolic type\cite{Pena2} and describes the
irreversible time evolution of the (real) amplitudes $\psi _{-}$ and $\psi
_{+}$.

If $\lambda=+1$, then equation (\ref{37}) becomes 
\begin{equation}
\label{40} \mp 2imD_0\partial\psi_{\pm}/\partial t=
-2mD_0^2\nabla^2\psi_{\pm}+V\psi_{\pm} 
\end{equation}
having as its solution 
\begin{equation}
\label{41} \psi_{\pm}=e^{R\mp iS}=\rho^{1/2}e^{\mp iS}\mbox{ onde }%
\rho=\psi_+\psi_-. 
\end{equation}
This is an hyperbolic type equation\cite{Pena2} and describes the reversible
evolution of the (complex) amplitudes $\psi_-=\psi$ and $\psi_+=\psi^{\dag}$.

If we put into equation (\ref{40}) 
\begin{equation}
\label{42} D_0=\frac{\hbar}{2m}, 
\end{equation}
where $\hbar$ is Planck's constant, we finally get the Schr\"odinger
equation.

With the definition 
\begin{equation}
\label{43} \rho=\psi^{\dag}\psi=e^{2R}, 
\end{equation}
we have, because of the second relation in (\ref{36}), 
\begin{equation}
\label{44} {\bf u}=D_0\frac{\nabla\rho}{\rho} \mbox{ , } D_0=E_t\frac{%
[(\delta {\bf x}(\Delta t))^2]}{2\Delta t}=\frac{\hbar}{2m}. 
\end{equation}
In the same way 
\begin{equation}
\label{45} {\bf f}={\bf f}_0+2m{\bf a}_s. 
\end{equation}

The results (\ref{44}) and (\ref{45}) will be very important for the
comparison between this derivation and the one based upon the infinitesimal
Wigner-Moyal transformation, to be presented in the next section.

The process of derivation of the equation related with Brownian movement and
the one related with the quantum formalism doesn't leave any doubt about the
irreducibility of one type of phenomenon into the other. Indeed, since the
very beginning, we have said that the quantum mechanical process has to be
understood as one where there is no room for friction, which distinguishes
it from the usual Brownian process.

\section{Second Derivation}

Now we will present again the mathematical demonstration of the
Schr\"ondiger equation from the Liouville equation and Newton's laws using
the infinitesimal Wigner-Moyal transformation, which has been our approach
since the first paper of this series. This will be done for the commodity of
references to be made latter on.

We begin with the three basic axioms

\begin{itemize}
\item  Newtonian mechanics is valid for all the particles composing the
systems of a given {\it ensemble};

\item  For an {\it ensemble} of isolated systems, the joint probability
density function is a conserved quantity 
\begin{equation}
\label{46}\frac{dF({\bf x},{\bf p};t)}{dt}=0;
\end{equation}

\item  The infinitesimal Wigner-Moyal transformation, defined as 
\begin{equation}
\label{47}\rho \left( {\bf x}+\frac{\delta {\bf x}}2,{\bf x}-\frac{\delta 
{\bf x}}2;t\right) =\int {F({\bf x},p;t)\exp \left( i\frac{{\bf p}\cdot
\delta {\bf x}}\hbar \right) d{\bf p}},
\end{equation}
may be applied to the problem.
\end{itemize}

We may write the Liouville equation as 
\begin{equation}
\label{48} \frac{dF}{dt}=\frac{\partial F}{\partial t}+\frac{{\bf p}}{m}%
\cdot \frac{\partial F}{\partial {\bf x}}-\frac{\partial V}{\partial {\bf x}}
\cdot\frac{\partial F}{\partial {\bf p}}=0, 
\end{equation}
where we have already used Newton's equation, valid according to the first
postulate.

With the characteristic function given by (\ref{47}) we may multiply the
Liouville equation by the exponential factor there defined and integrate
term by term in the momenta to get the equation 
\begin{equation}
\label{49}-i\hbar \frac{\partial \rho }{\partial t}-\frac{\hbar ^2}m\frac{%
\partial ^2\rho }{\partial {\bf x}\delta {\bf x}}+\delta {\bf x}\cdot \frac{%
\partial V}{\partial {\bf x}}\rho =0.
\end{equation}

Now, making the {\it ansatz} 
\begin{equation}
\label{50}\rho ({\bf x},\delta {\bf x};t)=\psi ^{\dag }({\bf x}-\delta {\bf x%
}/2;t)\psi ({\bf x}+\delta {\bf x}/2;t),
\end{equation}
using the infinitesimal character of the displacement $\delta {\bf x}$ and
writing the amplitudes as $\psi =exp(R+iS)$, with $R$ and $S$ as in the
previous section, we finally get the amplitude equation given by 
\begin{equation}
\label{51}-i\hbar \frac{\partial \psi }{\partial t}-\frac{\hbar ^2}{2m}\frac{%
\partial ^2\psi }{\partial {\bf x}^2}+V({\bf x})\psi =0,
\end{equation}
which is nothing but Schr\"odinger equation for the amplitude.

It is interesting to stress again that, as was shown in paper 1 of this
series\cite{eu1}, the equation (\ref{49}) has no dispersion associated to it
in a necessary manner, whereas for equation (\ref{51}) the Heisenberg
relations are necessarily valid. This result will be of some relevance in
what follows.

\section{Comparison Between the Derivations}

Now that both derivations were presented it remains the question about what
connection they maintain with each other. The unraveling of such a
connection may be of great importance to clarify some of the significances
of the quantities involved in the formalism, and also their place in the
underlying epistemology.

To find this connection, consider the transformation given in (\ref{1}) 
\begin{equation}
\label{66}\rho ({\bf x},\delta {\bf x};t)=\int_{-\infty }^{+\infty }e^{i{\bf %
p}\cdot \delta {\bf x}/\hbar }F({\bf x},{\bf p};t)d{\bf p},
\end{equation}
where $F({\bf x},{\bf p};t)$ is the joint probability density defined over
phase-space and $\rho ({\bf x},\delta {\bf x};t)$ is the corresponding
characteristic function.

We may expand the exponential with respect to the infinitesimal displacement 
$\delta {\bf x}$ and integrate in the momentum to get 
\begin{equation}
\label{67} \rho({\bf x},\delta {\bf x};t)=\rho({\bf x})\left[1+ \frac{i<{\bf %
p}>_F\cdot\delta {\bf x}}{\hbar}- \frac{<p^2>_F(\delta {\bf x})^2}{2\hbar^2}%
+O[(\delta {\bf x})^3]\right], 
\end{equation}
where $\langle\rangle_F$ represents the mean taken using the function $F(%
{\bf x},{\bf p};t)$.

Inserting this result into the equation for the characteristic function (\ref
{49}) and equating to zero the real and imaginary parts (or the terms in
zeroth and first order in the infinitesimal displacement) we get the set of
equations 
\begin{equation}
\label{68} \left\{ 
\begin{array}{l}
\partial\rho_0/\partial t+ 
\frac{<{\bf p}>_F}{m}\cdot\nabla\rho_0=0 \\ <{\bf p}>_F{\partial\rho_0}/{%
\partial t}+\frac{<p^2>_F}{m}\nabla\rho_0 +\rho_0\nabla V=0 
\end{array}
\right. , 
\end{equation}
with $\rho_0=\rho({\bf x},t)$ as in the second section, equation (\ref{43}).

Now using the first of these equations into the second, we find 
\begin{equation}
\label{69}\frac{(\Delta _F{\bf p})^2}m\nabla \rho _0+\rho _0\nabla V=0,
\end{equation}
where 
\begin{equation}
\label{70}(\Delta _F{\bf p})^2=<{\bf p}^2>_F-<{\bf p}>_F^2.
\end{equation}
If we make the identification 
\begin{equation}
\label{71}{\bf p}=m\frac{\delta {\bf x}}{\Delta t},
\end{equation}
then equation (\ref{69}) may be rewritten as 
\begin{equation}
\label{72}m\frac{(\Delta _t(\delta {\bf x}))^2}{\Delta t}\nabla \rho _0+\rho
_0\nabla V\Delta t=0,
\end{equation}
which may be understood, when we compare it with the results (\ref{44}) and (%
\ref{45}), as representing the equilibrium between the external force and
the stochastic component given by 
\begin{equation}
\label{73}{\bf f}_s=2m{\bf a}_s=2m\frac{{\bf u}}{\Delta t}=2m\frac{(\Delta
_t(\delta {\bf x}))^2}{2(\Delta t)^2}\nabla \rho _0,
\end{equation}
and coming from the fluctuations, where 
\begin{equation}
\label{74}(\Delta _t(\delta {\bf x}))^2=<(\delta {\bf x})^2>_t-<(\delta {\bf %
x})>_t^2,
\end{equation}
with the mean $<>_t$ defined as in (\ref{2}).

This is precisely the result one would expect from a stochastic approach to
the problem, as we have seen in the second section, and allows us to make
the connection between the two apparently distinct manners of deriving
quantum mechanics. This, in turn, gives us the possibility of studying their
respective ontological values.

The key to such a connection are the equations (\ref{70}) and (\ref{71}),
where the symbols of one approach (${\bf p}$ and $\delta {\bf x}$) are
written in terms of the symbols of the other ($\delta {\bf x}$ and $\Delta t$%
).

In the following subsection the relations between these symbols will be
finally made clear.

\subsection{Theoretical Connections}

We begin this section by first noting that the connection between the
momenta dispersion and the one related with the infinitesimal displacements
implies in fixing a minimum value for the time variation $\Delta t$. Indeed, 
\begin{equation}
\label{75} (\Delta_F {\bf p})^2=m^2\frac{(\Delta_t(\delta{\bf x}))^2}{%
(\Delta t)^2}, 
\end{equation}
where, using the expression 
\begin{equation}
\nu=D_0=\frac{(\Delta_t(\delta{\bf x}))^2}{2(\Delta t)}=\frac{\hbar}{2m}, 
\end{equation}
we get 
\begin{equation}
\label{76} \Delta t=\frac{m\hbar}{(\Delta_F {\bf p})^2}. 
\end{equation}

This last relation may be cast into a more familiar format. To show this we
write (\ref{76}) as 
\begin{equation}
\label{77}\Delta t=\frac \hbar 2\frac{2m}{(\Delta _F{\bf p})^2},
\end{equation}
and note that the last term in the right-hand side may be identified with a
dispersion in the kinetic energy $(\Delta _FE_k)$, then we get 
\begin{equation}
\label{78}(\Delta t)(\Delta _FE_k)=\frac \hbar 2,
\end{equation}
showing that the limitation in the time variation obeys a dispersion
relation (by its minimum value) with the kinetic energy. Since we must have
for the dispersion in the potential energy $(\Delta _FV)\ge 0$, the
dispersion in the total energy may be written as 
\begin{equation}
\label{79}(\Delta _FE)=(\Delta _FE_k)+(\Delta _FV),
\end{equation}
and we find, immediately 
\begin{equation}
\label{80}(\Delta t)(\Delta _FE)\ge \frac \hbar 2,
\end{equation}
as expected.

This result may, in fact, be considered as a demonstration of the validity
of the time-energy dispersion relation, by means of the concepts apprehended
from the stochastic approach. Note, however, that, while $(\Delta _FE)$ is a
dispersion, $\Delta t$ is not, being simply a variation. This was expected
since, as we know, the time $t$ is only a parameter in the quantum formalism
(has no operator associated to it) and is not subjected to dispersions.

What the stochastic approach shows is that, even not being related to
dispersions, its variation is fixed by a minimum value.

Since the time variation $\Delta t$ has a minimum value {\it distinct from
zero} the evolution of the {\it ensemble} in the interval $[t-\Delta t,t]$
may be distinct from its evolution in the next time interval $[t,t+\Delta t]$%
\cite{Pena2}. This fact is the very reason for the stochastic behavior to
appear in the process $\delta {\bf x}$.

This result establishes the relation between the two formal derivations of
the Schr\"odinger equation. We may now go to the conclusions.

\section{Conclusions}

Returning to equation (\ref{69}), we may argue that, while using the
characteristic function $\rho ({\bf x},\delta {\bf x};t)$ we had not imposed
upon the dispersions a value necessarily distinct from zero\cite{eu1}. In
this case, we still have 
\begin{equation}
\label{81}(\Delta _F{\bf p})=0,
\end{equation}
and the minimum value for the dispersion is introduced into the theory when
we write the characteristic function as the product 
\begin{equation}
\label{82}\rho ({\bf x},\delta {\bf x};t)=\psi ^{\dag }({\bf x}-\delta {\bf x%
}/2;t)\psi ({\bf x}+\delta {\bf x}/2;t),
\end{equation}
and find Schr\"ondiger equation for the amplitudes, as was already
demonstrated.

It becomes clear thus that the stochastic behavior of the theory is fixed
only at this latter stage, for if we have $(\Delta _F{\bf p})$, we still
keep ourselves within a purely newtonian-liouvillian description.

Therefore, the stochastic behavior is introduced by the characteristic
function decomposition as the amplitude product described above. We have to
remember that, according to paper 11 of this series\cite{eu11}, this
decomposition of the characteristic function is possible only if we may
decompose the probability density $F$ as 
\begin{equation}
\label{83}F({\bf x},{\bf p};t)=\int \phi ^{\dag }({\bf x},2{\bf p}-{\bf p}%
^{\prime };t)\phi ({\bf x},{\bf p}^{\prime };t)d{\bf p}^{\prime }
\end{equation}
when the amplitudes $\phi $, defined upon phase-space, are related with the $%
\psi $ amplitudes by 
\begin{equation}
\label{84}\psi ({\bf x}+\delta {\bf x}/2;t)=\int e^{i{\bf p}\cdot \delta 
{\bf x}/2\hbar }\phi ({\bf x},{\bf p};t)d{\bf p},
\end{equation}
as was already shown\cite{eu11}.

Such a decomposition in the probability density $F({\bf x},{\bf p};t)$
necessarily implies its non positive-definite character for the excited
states. Indeed, for these states, the $\psi$ amplitudes, and in general the $%
\phi$ amplitudes, have nodes (are `wave functions') and so, assume negative
values over some regions; being $F$ an autocorrelation function of these
amplitudes (and $\rho$ its convolution) it becomes clear that $F$ will
assume, in general, negative values for these states. Note, however, that
for the ground state there will be no nodes in the amplitudes, and these
amplitudes will be positive-definite. In this cases the function $F$ will
also be positive-definite.

This last observation may be a clue to understand the impossibility of
having the joint probability density function $F$ positive-definite for the
excited states. Indeed, these states are not, strictly speaking, stationary,
since they have a finite mean life-time associated to them. Only the ground
state is to be considered as really stationary. In this case, since the
stationary property of the states is related with the possibility of having
an equilibrium situation between the external force and the stochastic
one---equation (\ref{69})---we may expect that, for these states, this force
balance will be no longer valid, and equation (\ref{69}) will have only an
approximate validity.

In this sense, we may make the conjecture that the negative behavior of the
probability density for the excited states is reflecting exactly this fact.
The mathematical demonstration of this property, however, will be left to
another paper.

Now it remains to make some comments about the ontological aspects revealed
by the formal developments we have dealt with. To begin with, it is
important to stress again that the stochastic character of the theory
appears as a by-product of the momentum dispersion.

In a general problem we have the Heisenberg relation given by expression (%
\ref{80}). In the specific harmonic oscillator problem, for example, we have 
\begin{equation}
\label{85}(\Delta _FE)=\frac{\hbar \omega }2, 
\end{equation}
where $\omega $ is the frequency of oscillation characteristic of the
problem. For this problem, relation (\ref{80}) gives 
\begin{equation}
\label{86}\Delta t\ge \frac 1\omega =\frac T{2\pi }, 
\end{equation}
where $T$ is the characteristic period of the problem. The minimum time
interval to be considered for the stochastic formalism to be still
applicable is of the order of the movement period. Hence, it is not
unexpected that the energy dispersion makes it impossible to take for
granted in which specific state the oscillator is\cite{eu9}, for 
\begin{equation}
\label{87}E=(n+1/2)\hbar \omega \pm (\Delta _FE)=(n+1/2)\hbar \omega \pm
\hbar \omega /2. 
\end{equation}

It is also interesting that, for this specific example, the minimum time
interval $\Delta t$ is not really small compared with the characteristic
time of the system, as required in the development (\ref{2}). In fact, this
interval is of the same order or greater to this characteristic time. This
can be considered another argument to explain the result (\ref{87}).

It is important to stress that, being $t$ a parameter, we may, from the
experimental point of view, work with times much smaller than those allowed
by the equation (\ref{86}). In this case, we expect that the quantum
formalism, now seen as a stochastic theory, will be no longer adequate to
describe the problem.

Finally, we would like to stress that, as with the Brownian movement, the
fluctuations are a mathematical artifice introduced to simulate the particle
interactions with the environment, since the exact description of the
interaction may not be possible. Hence, in the Brownian movement, for
example, the fluctuations are caused by the collisions that, rigorously
speaking, obey Newton's equation, but whose exact consideration is
despairingly prohibited.

This is exactly our point of view.

\end{document}